\documentstyle[12pt]{article}

\def\alwaysmath#1{{\ifmmode{#1}\else{$#1$}\fi}}
\def\he#1{\hbox{\alwaysmath{{}^{#1}}{\rm He}}}

\def\li#1{\hbox{\alwaysmath{{}^{#1}}{\rm Li}}}
\def\b1#1{\hbox{$^{1#1}{\rm B}$}}
\def\c1#1{\hbox{$^{1#1}{\rm C}$}}
\def\n1#1{\hbox{$^{1#1}{\rm N}$}}
\def\o1#1{\hbox{$^{1#1}{\rm O}$}}
\def\be#1{\hbox{\alwaysmath{{}^{#1}}{\rm Be}}}

\def\minf{${\,m_{\rm inf}}$}

\def\etal{{\it et al.}~}
\def\beginapjbib{\begingroup \section*{\large \bf References}
   \parskip=.5ex plus 1.0pt
   \def\bibitem{\par \noindent \hangindent\parindent
      \hangafter=1}}
\def\endapjbib{\par \endgroup}

\def\la{~\mbox{\raisebox{-.7ex}{$\stackrel{<}{\sim}$}}~}
\def\ga{~\mbox{\raisebox{-.7ex}{$\stackrel{>}{\sim}$}}~}
\def\beq{\begin{equation}}
\def\eeq{\end{equation}}

\begin{document}

\begin{titlepage}

\pagestyle{empty}
\baselineskip=21pt

\rightline{UMN-TH-1415/95}
\rightline{astro-ph/9512034}
\rightline{submitted to {\it The Astrophysical Journal}}
\vskip .2in
\begin{center}
{\large{\bf LiBeB PRODUCTION BY NUCLEI AND NEUTRINOS}} \\
\vskip .1in

Elisabeth Vangioni-Flam$^1$, Michel Cass\'{e}$^2$, Brian D. Fields$^{1,4}$
and Keith A. Olive$^3$

$^1${\it Institut d'Astrophysique  de Paris, CNRS \\
98 bis bd Arago 75014 Paris,
France}

$^2${\it Service d'Astrophysique/DSM/DAPNIA/CEA \\
Saclay, France}

$^3${\it School of Physics and Astronomy, University of Minnesota \\
Minneapolis, MN 55455, USA}

$^4${\it Department of Physics, University of Notre Dame \\
Notre Dame, IN 46556}

\end{center}
\vskip .5in
\centerline{ {\bf Abstract} }
\baselineskip=18pt

The production of LiBeB isotopes by nuclear and neutrino spallation
are compared in the framework of galactic evolutionary models.  As
motivated by $\gamma$-ray observations of Orion, different possible
sources of low-energy C and O nuclei are considered, such as supernovae
of various masses and WC stars. We confirm that the low energy nuclei
(LEN), injected in molecular clouds by stellar winds and type II
supernovae originating from the most massive progenitors,
can very
naturally reproduce the observed Be and B evolution in the early galaxy (halo
phase). Assuming the global importance of the LEN
component, we compute upper and lower bounds
to the neutrino process contribution corresponding to limiting cases of
LEN particle spectra. A consistent solution is found with a spectrum
of the kind proposed by Ramaty \etal (1995a,b), e.g. flat up to $E_c=30$ MeV/n
and decreasing abruptly above. This solution fulfills the challenge
of explaining
at the same time the general Be and B evolution, and
their solar system abundances
without overproducing \li7 at very low metallicities, and the meteoritic
\b11/\b10 ratio. In this case, neutrino spallation is constrained
to play a limited role in the genesis of the solar system \b11.
Galactic cosmic rays (GCR)
become operative late in the evolution of the disk ([Fe/H]$>$-1), but
their contribution to the solar abundances of \be9, \b10 and \b11 is not
dominant (35\%, 30\% and 20\% respectively). Thus, with this
LEN spectrum, GCR are {\it not}\
the main source of \be9 and B in the Galaxy.
The most favorable case for neutrinos, (adopting the same kind of spectrum)
has $E_c=20$ MeV/n.  Even in this case,
the neutrino yields of Woosley and Weaver (1995) must to be reduced by a
factor of 5 to avoid \b11 overproduction.
Furthermore, this solution leads to a high B/Be ratio at
[Fe/H]=-2, difficult to reconcile with the observations,
unless specifically non-local thermodynamic equilibrium
corrections to the boron abundance are large.
On the other hand, if neutrino spallation does play an important role
in the production of galactic \b11, then LEN processes are relegated to a more
local phenomenon.  However, in this case, unless neutrino spallation
can also produce \be9 (and to some extent \b10 and \li6), a new source of
primary \be9 must be found.

\noindent
\end{titlepage}

\baselineskip=18pt

\section{Introduction}

Since the 1970's it has been known that spallation reactions between
high-energy nuclei in the galactic cosmic radiation (GCR) and the
interstellar medium are an important source of Li, Be, and B (LiBeB)
isotopes.  However, this mechanism is known to be incomplete, as
indicated by the well-established failure of GCR production to
reproduce the \li7 abundance in Population I stars, the meteoritic \b11/\b10
isotopic ratio,
and as shown more recently by the linear (rather than quadratic)
proportionality between Be (and B) and Fe in the early Galaxy
(Population II).  Thus there is a need for an additional source of
LiBeB.  Partly in response to this need, two new sources of LiBeB, and
particularly \b11 have been recently proposed:

\begin{enumerate}
\item Synthesis of
\li7 and \b11 by neutrino spallation in the helium and
carbon shells of supernovae (Woosley \etal 1990, Olive \etal 1994,
Timmes \etal 1995)--the so-called
``$\nu$-process."

\item Breakup of low energy nuclei (LEN) injected in
molecular clouds of the kind detected in Orion through their gamma ray
line emission (Bloemen \etal 1994, Cass\'{e} \etal 1995, Ramaty \etal
1995a,b,1996, Fields \etal 1995a).
\end{enumerate}

With the addition of these processes, spallation now depends on three
sources related to high energy ($>$ 500 MeV/n) and low energy ($<$
30 MeV/n) nuclei, and to neutrinos ($<$ 8 MeV).  The effect of the
$\nu$-process on LiBeB evolution has been considered by Olive \etal
(1994).  These authors followed Prantzos \etal (1993) in assuming that
GCR were more efficiently confined in the early galaxy. By itself,
such an ``overconfinement" model reproduced crudely the Be and
B evolutions (but not perfectly, see Tayler 1995). The most recent data confirm
the linearity of both the B and \be9 {\it vs}.\ Fe correlations
(Duncan \etal 1995;
Rebull \etal 1995).
As such they are indeed difficult to reconcile with the overconfinement model,
especially at very low metallicity where the calculated slope is 2.
Moreover, the \b11/\b10 ratio always remains unavoidably
close to 2.5, the usual GCR value. To obtain a solar boron isotopic
ratio of \b11/\b10=$4.05 \pm 0.05$ {
(Shima and Honda 1962, Chaussidon \& Robert 1995)},
Olive \etal (1994) introduced neutrino spallation, which
enhances \b11 (and not \b10) and so can give the right boron isotopic
ratio in the solar system. Since this process produces negligible amounts
 of \be9, one
has to invoke a separate production of this isotope by nuclear spallation.

Recently, the $\nu$-process yields have been updated (Woosley \&
Weaver 1995), and more importantly, the Bloemen \etal (1994) discovery
of unexpected $\gamma$-ray emission from Orion has provided strong
evidence for a LEN component.  The Orion $\gamma$-radiation is consistent
with line emission from \c12* and \o16* and as such can only be
plausibly explained by a large flux of low-energy nuclei enriched
in C and O. Given this observational
evidence for LEN, and its ability to copiously produce LiBeB, the
combined effect with neutrino spallation should be reconsidered.  In
this paper we will explore constraints on the relative contribution of
all three galactic LiBeB sources: GCR, LEN, and the $\nu$- process.
We will show that one can significantly constrain the relative
contribution of the different sources by exploiting the particularities
of each process and the interplay of these components in LiBeB
evolution.

The paper is organized logically as follows: we first introduce the three
components of LiBeB production and discuss the available free
parameters for each.  In particular, since the origin of the LEN is
still debated, as are the corresponding LEN source abundances, we
introduce various C and O enriched compositions related to stellar
winds and supernovae of different masses.  Next, the parameters for
the different production mechanisms are constrained by comparing to
the available observational and solar system data on LiBeB evolution.  These
data consist of the history of the elemental abundances (observed as a
function of metallicity [Fe/H]) and of the solar system isotopic
abundances and their ratios. Specifically, the adopted solutions
are bound by observational requirements: 1) to avoid Li overproduction
at low metallicity (i.e. preserve the Spite plateau, e.g. Spite and
Spite 1993); 2) to reproduce the quasi linear relationship between
Be/H and B/H {\it vs}.\ [Fe/H], at least up to [Fe/H] $<$ -1;
3) to reproduce the solar Be/H
abundance; 4) to generate a \b11/\b10 ratio such that, after mixing
with the GCR products (\b11/\b10 = 2.5) and the neutrino induced \b11
we get the observed meteoritic value of $\b11/\b10=4.05 \pm { 0.05}$
({Chaussidon and Robert 1995}).

We evaluate the possible models, showing that each process contributes
in a unique way to LiBeB evolution, and that the different behaviors
of the proposed mechanisms allow only a restricted set of possible
combinations of them that still fit the data.  In particular, we will
show that if the LEN component produces Be and B in the early Galaxy
as observed in Pop II stars, then this mechanism remains today the
dominant source of LiBeB, contrary to what has been previously thought.
Also, as a continuation of the work of Olive \etal (1994), we
determine the maximum contribution of the neutrino process to the
LiBeB production in the new context in which LEN are considered.

Two important assumptions should be made
explicit here.  (1) We take the LEN component, observed to exist in
Orion, to be ubiquitous to all star-forming regions throughout the
Galaxy's history.  Also, (2) we determine the LEN irradiation time such
that the particle
bombardment in these regions gives the
LiBeB abundances observed in extreme Pop II stars.  As we will argue,
the required exposure times
($\la 10^5$ yr) are  reasonable.

A word about the chemical evolution
models is in order.  We adopt the same standard galactic evolutionary
model as in Olive \etal (1994) and Vangioni-Flam \etal (1995).  Our
analysis is restricted to a simple closed box model including
production of light elements by GCR, LEN and neutrinos, and their
destruction in stars, as in Olive \etal (1994).  We take into account
the new neutrino yields released by Woosley and Weaver (1995) and
recent work on the low energy nuclei injected in active star
forming regions by Cass\'{e} \etal (1995), Vangioni-Flam \etal (1995)
and Ramaty \etal (1995a,b, 1996), analyzing in finer details SN of different
masses.

\section{The role of the  different components}

\subsection{LEN}
\label{sec:len}

The nuclear lines observed in Orion (Bloemen \etal 1994) must be
excited via low-energy nuclear interactions; the $\gamma$-ray emission
thus provides the first direct evidence of a LEN component.
As the astrophysical source of LEN is uncertain, we will try
three different models (\S\S 3.1--3.3).  The first of these models
to have been considered is the one most studied in the literature, and
the one that does the best job of giving LiBeB evolution; we will thus
focus on this model for the moment.
In this scenario,
 the production rates of the LiBeB isotopes have been
determined adopting a simple source spectrum of the kind
$N(E) = const$ up to $E=E_c$,
and $N(E) \propto E^{-n}$ above $E_c$.
We follow Cass\'{e} \etal (1995), and Ramaty \etal (1995a,b) in assuming
all nuclear species to have the same source spectrum,
which we propagate down to
thermalization in the cloud.
We assume that Orion is representative of all the active star forming regions
in the galaxy at all times; consequently the production rates
need to be multiplied by the
irradiation or exposure time $\tau$ to get the corresponding yields.
The LEN mechanism is thus determined by a source composition
(or---if one assumes the sources to come from a variety of supernovae---the
lower limit $m_{\rm inf}$ of this spectrum), the exposure time $\tau$,
and the spectral parameters $n$ and $E_c$.
(A summary of the production mechanisms, the
free parameters in each, and how these are fit is found in
Tables 1 and 2.)

The LEN component is uniquely
capable of naturally producing
significant Be in the early Galaxy, and moreover it
naturally gives the observed approximately linear relation between Be
(as well as B) and Fe/H.
We will thus {\it require}\ that LEN reproduces the early galactic Be;
having done so, the yields of Li and B will depend on the particular
LEN particle spectrum chosen. Thus the details of the LEN contribution
(and thus of the GCR and $\nu$-process contributions) to LiBeB analysis
depend upon the spectrum assumed, and so some investigation of spectral
dependence is in order.

For the present work, however, rather than examining the full parameter space
of possible LEN spectra, i.e., of $(n,E_c)$ pairs, we will instead
focus on two interesting limiting cases.  The first is the case
emphasized in the literature so far, in which one searches for
spectra which give LEN nucleosynthesis yields consistent with all the
constraints discussed above. This will have the effect of minimizing the
need for contributions from the other sources, in particular from the
$\nu$-process.  The LEN spectrum for this case was first sought by
Cass\'e \etal (1995), who favored an $E_c \sim 10$ MeV.  Since then
it has been pointed out
(Ramaty \etal 1995a, b) that energetic considerations favor $E_c \sim 30$ MeV.
We will examine this more recent model, which has an index $n=9$.

Secondly, we will consider the case of the spectrum which has the
{\it lowest} possible contribution of \b11 given the Be production
(i.e., the lowest \b11/Be ratio) corresponding to $E_c=20$ MeV/n
(also with $n=9$).  This gives
the maximal room for a $\nu$-process contribution.  Thus this will set
an upper limit to the needed $\nu$-process yields for LEN-dominated
LiBeB evolution.

For a given spectrum, the remaining parameters for the LEN component
are the composition and the irradiation time $\tau$.
We will try different compositions, both of particular kinds of supernovae
(i.e., those with Wolf-Rayet progenitors), and of ensembles of supernovae
of different masses.
In our analysis we will adjust $\tau$
to fit the observed  rise of Be/H at low [Fe/H]. Note that B/H
in this metallicity regime is thus fixed, and becomes a prediction
of the model.  The contribution
of LEN at higher metallicity is judged by comparison with the observed
ratios.

\subsection{Galactic Cosmic Rays}

The GCR is treated classically (i.e., without the overconfinement
used in Prantzos \etal 1993).
The relative production of the different isotopes is
taken from the rates given by Read \& Viola (1984);
their intensity is taken
proportional to the supernova rate and is taken as a free parameter.  Since
the $\nu$-process
does not produce Be, for a given LEN production rate, the GCR
component must add whatever extra Be is necessary to obtain the solar
abundance.  Thus the GCR intensity is fixed to obtain the right Be/H at solar
birth, adding their contribution to the LEN one.  (Fitting the intensity is
equivalent to determining the present effective cosmic ray flux $F$;
the implications of this point are further discussed in Lemoine \etal 1995).
Having fixed the GCR contribution to give the correct solar Be abundance,
the abundances of \b10 and \b11 become predicitions of the model.

\subsection{Neutrinos}

Neutrino spallation is a source of \li7 and \b11 via the interactions of
neutrinos (predominantly $\nu_\mu$ and $\nu_\tau$) on  nuclei,
and of particular importance here on \he4 and \c12
(Woosley \etal 1990, Woosley and Weaver 1995 -
hereafter WW). The lithium and boron yields are quite
sensitive to the temperature of the
$\mu$ and $\tau$ neutrinos, in which there is a fair amount of
uncertainty (e.g. Janka \& Muller
1995). As a result, the overall yields of \li7 and \b11
have considerable uncertainties and
we will consider variations in the overall production level.  In
Olive \etal (1994),
$\nu$-process nucleosynthesis was incorporated into
a model of galactic chemical evolution with the primary purpose of augmenting
the low value for \b11/\b10 produced by standard GCR nucleosynthesis.
To correctly fit the observed ratio of 4, it was found that the yields of
Woosley \etal (1990) had to be tuned down by a factor of about 2 to avoid the
overproduction of \b11. Tuning down the $\nu$-process yields ensured as
well that the production of \li7 was insignificant, and did not affect
the Spite plateau.  However, the resulting ratio of B/Be was predicted
not only to be high, but also to be metallicity dependent
with B/Be increasing at low metallicity.  The recently non-LTE corrected
(Kiselman 1994; Kiselman \& Carlsson 1994)
B/Be ratio at low metallicity may lend support
to the enhanced production of boron at early times (Fields, Olive \&
Schramm, 1995b).

Recently, the $\nu$-process yields have been updated and now include
a metallicity dependence (WW).  We employ these new yields,
but recognizing the uncertainties in the input physics
(see \S \ref{sec:nudis}) we allow the magnitude of the yields to be
a relatively free parameter, scaling them by a factor $f$.
Figures 1 and 2 show that the full yields ($f=1$) lead to a marginal
overproduction of \li7 around [Fe/H]=-1; while \b11 production
by neutrinos is so efficient that no room is left for
other mechanisms. This case is excluded since other processes are necessary to
produce \be9 which consequently introduce their own \b11 contribution. If
we reduce the yields by a factor of 2 ($f=0.5$)
(full lines in the figures), we can allow for other mechanisms
contributing to the observed \b11 abundance.
Indeed, when the LEN processes are included to account for Population II \be9,
we will see that a further reduction in the $\nu$-process yields are
necessary.
Incidentally, note that at solar metallicity, the new yields
are about twice the old ones. Olive \etal (1994) had divided the old
yields by 2, which corresponds to a division by 4 of the new ones.
In the discussion below, $ f$ is adjusted to give the right \b11/\b10;
once the LEN and GCR contributions have been fixed to fit the Be
evolution, the B evolution in these models becomes a prediction.
As we will see in the next section, when all the components are taken
into account, the needed reduction in \b11 yields will depend
on the LEN spectrum adopted.

\section{Evolution driven by typical LEN progenitors}

The calculation includes sequentially the three components: LEN to
adjust the Be/H evolution, GCR to fit the solar abundance of
\be9 and possibly neutrinos to explain the meteoritic \b11/\b10 ratio.
This procedure will be applied in turn to the different LEN sources
proposed: massive stars (typically 60 M$_\odot$) (Ramaty \etal
1995a,b, Vangioni-Flam \etal 1995); the external layers of type Ic
supernovae (Fields \etal 1995); and ejecta of supernovae between
 15 and 100 M$_\odot$, given
high velocities in the explosion and then further reaccelerated
in the local medium.

\subsection{The most massive stars}

\label{sec:WC}

The WC model adopted by Ramaty \etal (1995a,b, 1996) is one of the best
LEN candidates. Moreover, including pre- and post-WC stages of
evolution (O, Of, WN, and explosion) or considering the explosion
of a massive star
of 60 M$_\odot$, not affected by mass loss, as in the early Galaxy
(Maeder 1992), we get results similar to the WC case (Vangioni-Flam
\etal 1995). So the WC case will serve as a reference for all stars of
$\sim 60$ M$_\odot$ on the main sequence, independent of their
metallicity and related mass loss.

As stated in \S \ref{sec:len}, we tried two limiting spectra.
For the $E_c = 30$ MeV/n spectrum,
the required irradiation time is
$\tau = 5 \times 10^4$ yr. A modest GCR contribution (35, 30 and 20\%
for \be9, \b10 and \b11), is sufficient to bring the calculated
\b11/\b10 ratio at solar age to its observed value of
$4.05 \pm {0.05}$, leaving virtually no
room for neutrino spallation. The \b11/\b10 ratio evolves from 4.8
(the LEN value) at
the beginning of galactic evolution, to 4 at solar birth
(fig.\ 3, full line).
The derived B/Be ratio, evolving from 26 and 20, also agrees
with observations at all metallicities (fig.\ 4, full line).
Accordingly, in this
case, the WW \b11 neutrino yields
are constrained to be reduced by a factor of at
least 10 ($f=0.1$) to avoid overproduction of this isotope.

The proposed solution, based on a WC composition and $E_c=30$ MeV/n seems
satisfactory since it explains at the same time the general \be9 and B
 evolution without overproducing \li7 in the early galaxy and the \b11/\b10
ratio in the solar system. However, a more involved solution, combining
 GCR, LEN and neutrinos cannot be a priori excluded.

To explore the strength of the constraints on the $\nu$-process
in our LEN-dominated scenario,
we employed a LEN spectrum tuned to minimize \b11 production, and thus
leave maximal room for a $\nu$-process contribution, while still
explaining the population II \be9 evolution.
Indeed, the \b11/\b10 production ratio {\it vs}.\ $E_c$ for a WC composition
(Ramaty \etal 1996) presents a minimum of about 3 at 20 MeV/n (compared to
about 5 at $E_c=30$ MeV/n at zero metallicity). Higher values of $E_c$ are
excluded since they lead to line widths inconsistent with the COMPTEL
data (Ramaty \etal 1995 a,b; Tatischeff \etal 1996).
 While this apparently excludes
an $E_c$ as high as 100 MeV/n (for which the $\nu$-process contribution
can be somewhat larger), it should be recalled that the line width data
of Bloemen \etal (1993) are subject to some uncertainty.

Taking spectra with $E_c=20 MeV/n$, fitting the lower bound of the early Be
evolution, and adjusting the cosmic ray
flux to get the solar value of \be9, we find an irradiation time of $ 5
\times 10^4$ yr
(this is equivalent to minimizing the the Orion contribution;
see fig.\ 5.)
In this case with $f=0.5$, we get a reasonable Be and B evolution, however
we also get $\b11/\b10 =7$ at the solar epoch which is far too large.

Therefore, we adopt $f=0.2$ which gives $ \b11/\b10 =4.5$, (fig.\ 3), which is
close to the observed value. Any further reduction would lead us to the first
case where neutrinos are negligible.  For \b11, the neutrino contribution
in this case is 40\%, the GCR one is 25\% and the Orion one is 35\%. Concerning
\be9, the respective contributions are 60\% for GCR and 40\% for Orion.
Finally, for \b10, we get 55\% for GCR and 45\% for Orion.

However, as shown in figure 4, the neutrino addition alters
maximally the B/Be ratio at [Fe/H] about -2. Below this value, the
Orion component
dominates since it is produced by massive (fast evolving )
stars which eject their C and O very  rapidly, whereas the neutrino component
is delayed since it arises from all stars more massive than 11 M$_\odot$.
Above [Fe/H] = -2, the B/Be ratio
decreases due to the growing importance of GCR. This bump in the
B/Be {\it vs}.\ [Fe/H]
curve provides a clear test to the operation of the neutrino mechanism
in conjunction to the LEN process.

The recent data of Duncan \etal (1995), lead to a B/Be ratio of about 10,
irrespective of [Fe/H]{}. Non-LTE corrections could enhance this ratio at low
metallicity (Kiselman 1994; Kiselman \& Carlsson 1994; Thevenin \etal 1996).
This correction, which is dependent on stellar parameters
such as surface temperature and gravity, may be as large as
a factor of 5 at [Fe/H]=-2.However, as shown in figure 4, there is a bump
in the B/Be evolution and the NLTE correction decreases
continuously with increasing metallicity. thus, the NLTE correction at [Fe/H]=
-2 should not be greater than the correction at [Fe/H]=-3. This may
 be indicative of an intrinsic inconsistency between the neutrino and LEN
processes.
A measurement of \b11/\b10
in stars at this metallicity (which is not out of reach of the HST; Duncan
1995, private communication) would be of considerable interest since the
predicted ratio is twice solar in this metallicity range, (fig.3).

The WC case posits
that only very massive stars contribute to the LEN component, i.e.,
the lower limit $m_{\rm inf}$
to the LEN sources is $m_{\rm inf}$ = 60M$_\odot$.
Can other supernovae contribute to the LEN component?  As the source
of these particles remains unclear, this question remains an open one,
but we can get an important hint by
noting the Li/Be ratio produced by lower mass supernovae.
For $m_{\rm inf} <$ 60M$_\odot$, one has Li/Be $\ga 100$;
thus an attempt to fit the Be abundance at low metallicities
will lead to a conflict with the Spite plateau.
The high Li/Be production ratio can be understood physically as being due
to the composition of the ejecta of lower-mass supernovae.  Yields
{}from these stars have large amounts of H and more importantly He, which
is very effective, via the $\alpha+\alpha$
reaction, in producing Li.  Thus it is
clear that to have an acceptable Li production, such stars must be
avoided and so $m_{\rm inf}$ must be fairly large; in fact we find the lowest
acceptable level is $m_{\rm inf}$ $\ge$ {35} M$_\odot$.
This conclusion holds for both of the proposed spectra.

\subsection{The  external layers of a SNIc }

If a supernova progenitor has
been subject to heavy mass loss, then
its outermost ejecta are propelled to high velocity
corresponding to several MeV/n
(Nomoto \etal 1995, Woosley, Langer and Weaver
1993). The direct impact of the fastest particles on the surrounding
medium should produce a copious amount of $\gamma$-ray lines of C and O,
as observed in Orion (Fields \etal 1995). This case is particularly
clean since there is no free parameter in the problem. The energy
spectrum and the number of fast nuclei of the different kinds are
derived directly from the calculated velocity and composition profiles
of the external part of the supernova. The spectrum is close to
a power law
with index 4 and a maximum energy $\sim 10$ MeV/n.
Here the irradiation time $\tau$ is not a free parameter, but
is identified
with the stopping time of the nuclei ejected imposed by ionization
losses in the cloud medium (i.e. a few $10^3$ yrs).
Since the exposure time is fixed,
we may not adjust it, as done in the previous section, to ensure
that the light element yields make a significant contribution to
LiBeB evolution; instead the yields are fixed.
Unfortunately, due to the shorter irradiation time compared to the
WC case, the LiBeB yields
obtained are of order $10^{-9}$ M$_\odot$ (compared to
typical $\nu$-process yields of $\sim 10^{-7}$ M$_\odot$)
and fall short of explaining the observations.

On the other hand,
this kind of process could serve as excellent injector to subsequent
acceleration.  In such a scenario, shock acceleration of the material
ejected by SN (as well as stellar winds) is required to get efficient
production of light elements by C-O rich ejecta
(see the following section ).
If further acceleration of SN and wind material does not
take place at all (which is unlikely, see e.g. Nath \& Biermann 1994;
Biermann 1995; Bykov \& Bloemen 1994), then to fit B in the early
galaxy demands the full contribution of neutrino production
($f=1$).
But in this case, \be9 is underproduced in the
early galactic times.
The overconfinement scenario (Prantzos \etal 1993)
being excluded since it leads to a slope of 2, a
complementary process of primary nature is required
(see e.g. Tayler, 1995), which would probably
bring
its own boron contribution. Consequently, once again, the neutrino contribution
would have to be reduced.

\subsection{Production by reaccelerated supernova ejecta}

We consider now the contribution of all supernovae (15--100 M$_\odot$;
Weaver \& Woosley 1993; Woosley, Langer \& Weaver 1993) assuming
(re)acceleration in their ejecta with the same spectrum the case
of the direct ejecta ($n=4$), but pushing the maximum energy up to
30 MeV/n instead of 10 MeV/n. This choice of spectrum parameters
does not contradict any observational constraint.
As with the WC case (\S \ref{sec:WC}), the composition and irradiation
(reacceleration) time remain free parameters.

Adopting these
spectral characteristics, we have varied the lower mass limit of the
contributing supernovae from $m_{\rm inf} = 15$ to
60 M$_\odot$. As expected, the
lower \minf, the higher is the average yields, and the shorter
$\tau$.  Further, $\tau$
varies between 500 yrs ($M>$15M$_\odot$) and 50 000 yrs,
($M>$ 60M$_\odot$).  The supernovae of lower mass
($M<$ 25M$_\odot$) are so efficient that they leave no room for the
$\nu$-process or GCR contributions,
which is quite unrealistic; thus we are again driven to favor
the massive stars alone as sources (M above 35 M$_\odot$).
For the mass spectrum
chosen in \S 3.1 ($M > 60$ M$_\odot$), the \b11/\b10 ratio is
about 3 instead of 4 and a slight neutrino contribution is allowed as in the
 previous case ($f=0.2$).
In this specific situation, an important part
of the solar
system lithium, if not all, can be produced by LEN in the
disk  due to the operation of the $\alpha + \alpha$ reaction.
Indeed, more \li6 and \li7 are produced than in the case of the most massive
stars
(M$>$60 M$_\odot$) since supernovae between 35 and 60 M$_\odot$
are more helium rich than
above this mass (Weaver \& Woosley 1993).

\subsection{Implications for the ${\bf \nu}$-Process}
\label{sec:nudis}

If the LEN contribution is a significant source of Be
in the early Galaxy,
then it necessarily produces copious B as well.
Thus to avoid B overproduction in early and solar epochs,
the $\nu$-process yields are required to be reduced from
their calculated values.  Is such a reduction plausible?
To be sure, supernovae are known to create a huge neutrino
flux, as observed in SN 1987A.  However, the details of
the neutrino spectrum are more model-dependent. The largest
uncertainties lie in the neutrino temperature (which controls
the neutrino flux) and the neutrino spallation cross-section.
Both sources carry at least a factor of two uncertainty (Woosley 1995,
private
communication).  The LEN spectrum with $E_c = 20$ MeV, and a reduction
in the $\nu$-process yields by a factor of five, is probably at the limit of
compatibility between these two mechanisms.
Note that if the $\nu$-process yields can be verified,
(e.g., via $^{19}$F observations (Timmes, Woosley \& Weaver 1995),
 or through a confirmation of the details of
the model parameters used to get the Li and B yields) then
there is a potential conflict with the LEN hypothesis. In this case, a new
source of primary \be9 would be required.

\section{Conclusion}

We have estimated the contribution of the different spallation
processes (nuclei at high and low energies and neutrinos at low
energy) to the production and evolution of the LiBeB isotopes.
Low energy nuclei are required to explain the Orion $\gamma$-ray observations,
and must have composition that is overabundant in C and O.
Examining specific sources, we find that
nuclei reflecting the composition of C and O rich stellar
ejecta (WC and/or massive undressed SN),
moderately accelerated in the
source vicinity by shock waves or turbulence, would play a significant role
in the story of Be and B.
Depending on the spectrum chosen, there is a varying amount of
room for the GCR and neutrino
processes; but regardless of the spectrum the low-energy nuclei
contribution is important.
This unique process is able to explain consistently the Be
and B evolution and solar system abundances (using the meteoritic B/H
ratio), and the meteoritic \b11/\b10 ratio, with the GCR playing
a secondary role, especially at low metallicity.
A more detailed work, involving updated GCR
production rates and a more refined analysis of the different
components will be presented in a forthcoming paper (Lemoine \etal
1995).

While WC-type sources provide particle compositions that
are both good candidates for Orion and which give good LiBeB evolution,
direct ejecta seem to do only the former; their LiBeB production
is negligible due to the short particle lifetime (ionization stopping time).
With reaccelerated ejecta,
the LEN component can again provide a key component in a good fit to LiBeB
evolution, but only for source compositions like those of massive
supernovae.  Thus we see that, at least for the cases we have studied,
massive progenitors for LEN particles are needed to provide
a good LiBeB evolution.  The details of such a selection effect
remain to be shown; our results provide a hint to model-builders.

Since we have to reduce ($E_c=20$ MeV/n) the neutrino yields
by a factor 5 even in the best case,
we fall short explaining Li in population I stars. In this case,
stellar sources are required as AGB stars (Abia, Isern and Canal 1993).
They should be rather long living (small masses) otherwise the Spite plateau
would be violated. Similarly, even if $\nu$-process nucleosynthesis is the
dominant mechanism in the production of \b11, the \b11/\b10 ratio
necessitates a reduction in the neutrino spallation yields and
would also preclude the $\nu$-process from explaining the population I
Li abundances (Olive \etal 1994).

In closing, it is worth noting that the solar B/H ratio is
still uncertain, varying from $2 \times 10^{-10}$ (within a factor of
2 for Pop I stars, Boesgaard and Heacox 1978) and $(7.5 \pm 0.6)
\times 10^{-10}$ in carbonaceous chondrites (Anders and Grevesse 1989),
to the photospheric value of $4 \times 10^{-10}$ (Khol
\etal 1977). As a byproduct of this work, we remark that in all the
fits the meteoritic B/H value is favored. The progress in this field
is now linked to better abundance determination.

\bigskip

 We are grateful to Tom Weaver, Stan Woosley,  and Frank
Timmes for access to the neutrino yields and the latter two for helpful
conversations. We thank Yvette Oberto and Roland Lehoucq for
permanent help.
This material was based on work supported by the North Atlantic Treaty
Organization under a grant awarded in 1994.
This work was supported in part by PICS n$^\circ$114,
``Origin
and evolution of the light elements," CNRS.
This work  was  also supported
in part by  DOE grant DE-FG02-94ER-40823.

\vskip 2 cm

\beginapjbib

\bibitem Abia, C., Isern, J. \& Canal, R. 1993, A\&A. 275, 96

\bibitem Anders, E. \& Grevesse, N. 1989, Geochim. et Cosmochim. Acta,
35, 197

\bibitem Biermann, P.L. 1995, in Cosmic Winds and
Heliospheres, ed.\  J.R. Jokipii, C.P. Sonnet \& M.S. Giampapa

\bibitem Bloemen H. \etal. 1994, A\&A, 281, L5

\bibitem Boesgaard, A.M. \& Heacox, H.D. 1978, Ap.J., 226, 888

\bibitem Boesgaard A. \& King J.R. 1993, AJ, 106, 2309

\bibitem Bykov, A. \& Bloemen, H. 1994, A\&A, 283, L1

\bibitem Cass\'{e}, M. Lehoucq, R. \& Vangioni-Flam, E. 1995, Nature,
373, 318

\bibitem Chaussidon, M. \& Robert, F. 1995, Nature, 374, 337

\bibitem Duncan D., Lambert D. \& Lemke M. 1992, ApJ, 401, 584

\bibitem Duncan, D. \etal.\ 1995, Maryland conference : Cosmic Abundances,
in press

\bibitem Fields, B.D., Cass\'{e} M. \& Vangioni-Flam, E., Nomoto K., 1995a,
ApJ, in press

\bibitem Fields, B.D., Olive, K.A., \& Schramm, D.N. 1995b, ApJ, 439, 854.

\bibitem Gilmore G., Gustafsson B., Edvardsson B. \& Nissen P.E. 1992,
Nature 357, 379

\bibitem Janka, H. Th.,Muller, E., 1995, A\&A, submitted

\bibitem Kiselman, D. 1994, A\&A, 286, 169

\bibitem Kiselman, D. \& Carlsson, M. 1994, in The Light Element Abundances,
ed.\ P. Crane, 372

\bibitem Kohl, J.L., Parkinson, W.H. \& Withbroe, N. 1977, ApJ, 212,
L101

\bibitem Lemoine, M., Vangioni-Flam, E. \& Cass\'e, M., 1995, in preparation

\bibitem Maeder A. 1992, A\&A, 264, 105

\bibitem Nath, B.B. \& Biermann, P.L. 1994, MNRAS, 270, L33

\bibitem Nomoto, K., Iwamoto,K. \& Suzuki, T. 1995, Physics Reports, in press

\bibitem Olive K.A., Prantzos N., Scully S. \& Vangioni-Flam E. 1994,
ApJ, 424, 666

\bibitem Prantzos, N., Cass\'{e},M. \& Vangioni-Flam,E. 1993, Ap.J.,
403, 630

\bibitem Ramaty R. , Kozlovsky B. \& Lingenfelter R.E., 1995a, ApJ,
438, L21

\bibitem Ramaty R., Kozlovsky B. \& Lingenfelter R.E. 1995b, in
Proceedings of the 17th Texas Symposium, New York Acad. Sci., in press

\bibitem Ramaty, R., Kozlovsky, B. \& Lingenfelter, R.E. 1996, ApJ, in press

\bibitem Read, S.M.,\& Viola,V.E., 1984, Atomic Data and Nuclear Data Tables,
31, 359

\bibitem Rebolo, R., Molaro, P., Beckman J.E. 1988, A\&A, 192, 192

\bibitem Rebull, L.M. \etal, 1995, Maryland conference: Cosmic Abundances,
in press

\bibitem Ryan, S., Bessell, M., Sutherland, R.\& Norris, J. 1990, ApJ, 348, L57

\bibitem Ryan S., Norris I., Bessel M. \& Deliyannis C. 1994, ApJ,
388, 184

\bibitem Shima, S.and Honda, W., 1962, J. Geophys. Res., 68, 2849

\bibitem Spite F. \& Spite M. 1993, "Origin and Evolution of the
Elements, ed. N. Prantzos, E. Vangioni-Flam \& M. Cass\'{e},
Cambridge University Press, 201

\bibitem Tatischeff, V. \etal 1996, in preparation

\bibitem Tayler, R.J. 1995, MNRAS, 273, 215

\bibitem Thevenin, F. \etal, 1996, in preparation

\bibitem Timmes, F.X., Woosley, S.E. \& Weaver, T.A. 1995, ApJ in
press

\bibitem Vangioni-Flam E., Cass\'{e} M. \& Ramaty, R. 1995, ApJ,
submitted

\bibitem Weaver T. M. \& Woosley S.E. 1993, Phys. Rep. 227, 65

\bibitem Woosley,S.E., Hartmann,D., Hoffman, R., Haxton, W. 1990, ApJ.,
356, 272

\bibitem Woosley, S. E., Langer, N. A., Weaver T. A. 1993, ApJ, 411,
823

\bibitem Woosley, S.E. \& Weaver, T.A. 1995, ApJ, submitted

\endapjbib

\newpage

\begin{table}[htb]
\label{tab:prod}
\caption{LiBeB Production Sites by Isotope}
\begin{center}
\begin{tabular}{cccccc}
\hline\hline
 & & & Isotope & & \\
Source & \li6 & \li7 & \be9 & \b10 &\b11 \\
\hline
LEN & yes & yes & yes & yes & yes \\
GCR & yes & yes & yes & yes & yes \\
$\nu$-process & no & yes & no & no & yes \\
\hline\hline
\end{tabular}
\end{center}
\end{table}

\begin{table}[htb]
\label{tab:param}
\caption{The Interplay of Source Parameters and Observables}
\begin{center}
\begin{tabular}{ccc}
\hline\hline
Source & Parameter & Fixed by \\
\hline
LEN: steady state & irradiation time $\tau$ & Be/H at [Fe/H] $\le$ -1 \\
LEN: direct ejecta & none & N/A \\
LEN: reaccelerated ejecta & lower mass limit \minf &  varies \\
GCR & total flux $F$ & solar Be/H \\
$\nu$-process & yield normalization $f$ & solar \b11/\b10 \\
\hline\hline
\end{tabular}
\end{center}
\end{table}

\newpage
\noindent{\bf{Figure Captions}}

\vskip.3truein

\begin{itemize}

\item[]
\begin{enumerate}
\item[]
\begin{enumerate}
\item[{\bf Figure 1:}] Evolution of \li7/H from neutrino
 spallation only.

The metallicity dependent yields from  Woosley
and Weaver (1995) are
multiplied by factors $f=1$ (dotted line), $f=0.5$ (full line).
Data points are
from Spite and
Spite (1993), Rebolo \etal (1988).

\item[{\bf Figure 2:}] Evolution of B/H from neutrino spallation
only.

Same as fig.\ 1.  Data points from Duncan \etal (1992) and
Kiselman \& Carlsson (1994).  The three values at solar [Fe/H],
from bottom to top
correspond to solar photospheric (Khol \etal 1977), present galactic
average (Boesgaard and Heacox 1978) and meteoritic (Anders and
Grevesse 1989).

\item[{\bf Figure 3:}] Evolution of \b11/\b10.

Full line: LEN ($E_c=30$ MeV/n)
+ GCR; dotted line: LEN ($E_c=20$ MeV/n) + GCR + neutrinos ($f=0.2$).

\item[{\bf Figure 4:}] Evolution of B/Be.

 same as figure 3.

\item[{\bf Figure 5 :}] Evolution of Be and B.

LEN ($E_c=20$ MeV/n) + GCR +
neutrinos ($f=0.2$). Data points are from Ryan \etal (1990),
Gilmore \etal (1992),
Ryan \etal (1994) and Boesgaard and King (1993) (Be), Duncan \etal (1992)
and Kiselman \& Carlsson (1994) (B).

\end{enumerate}
\end{enumerate}
\end{itemize}

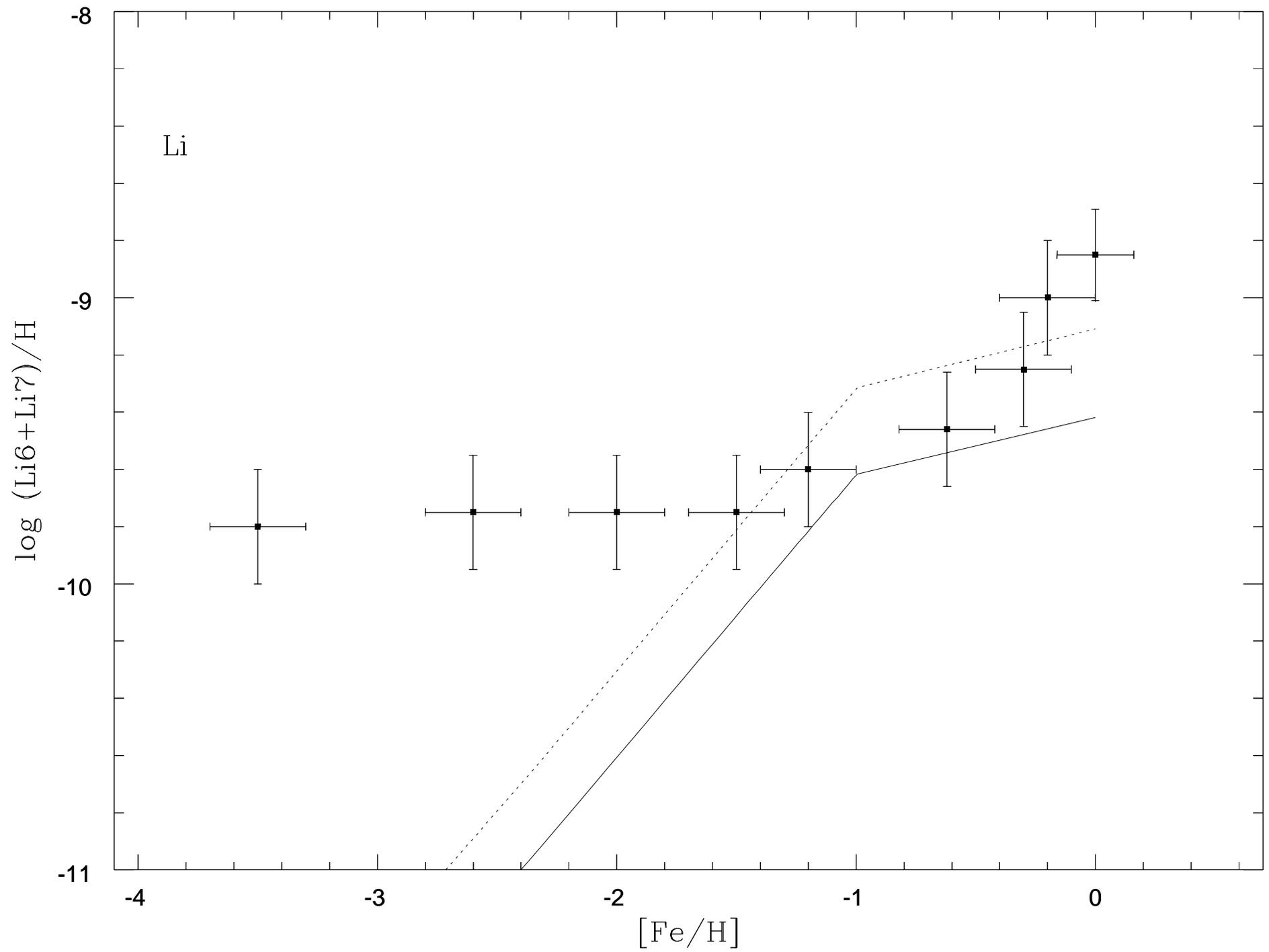
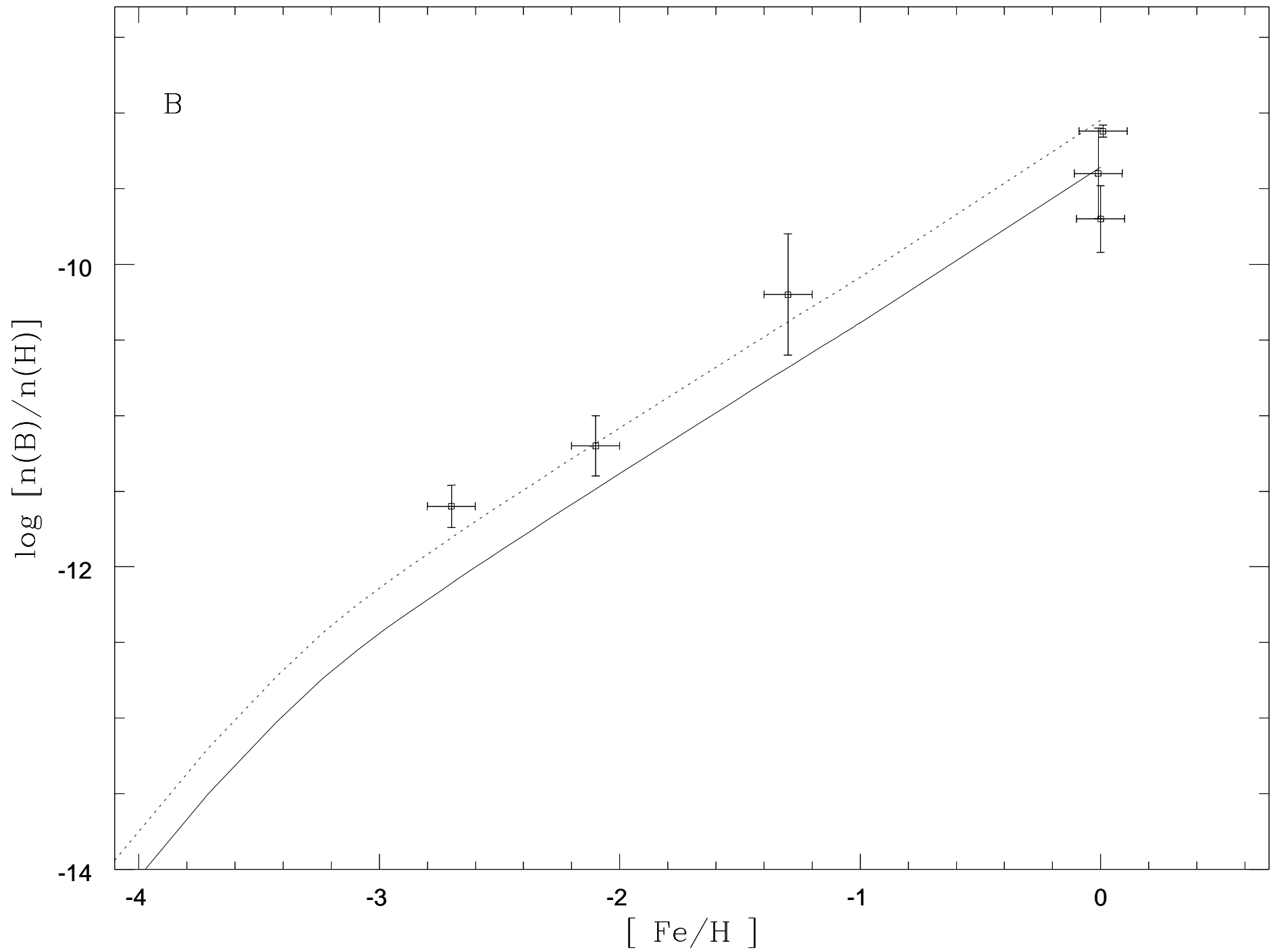
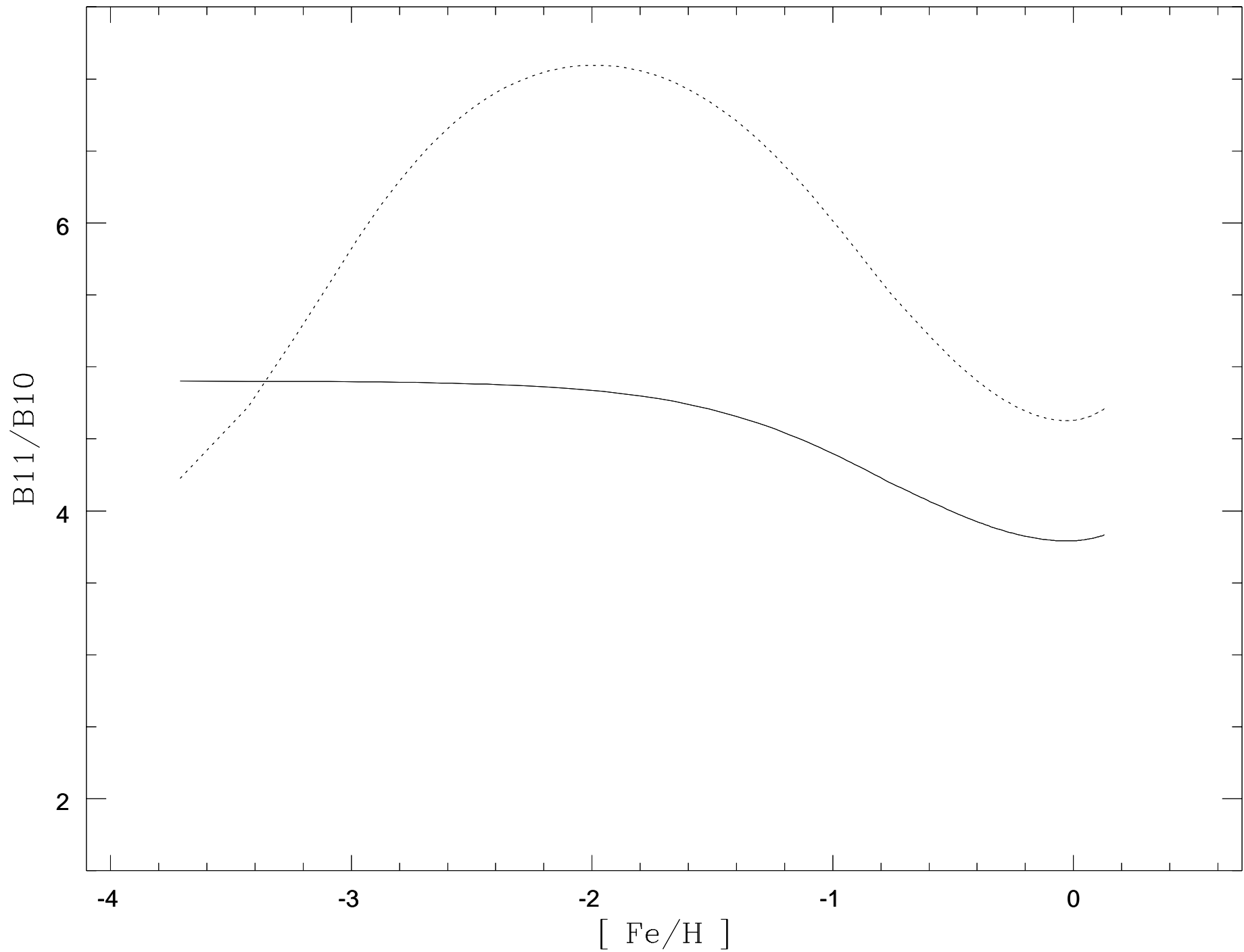
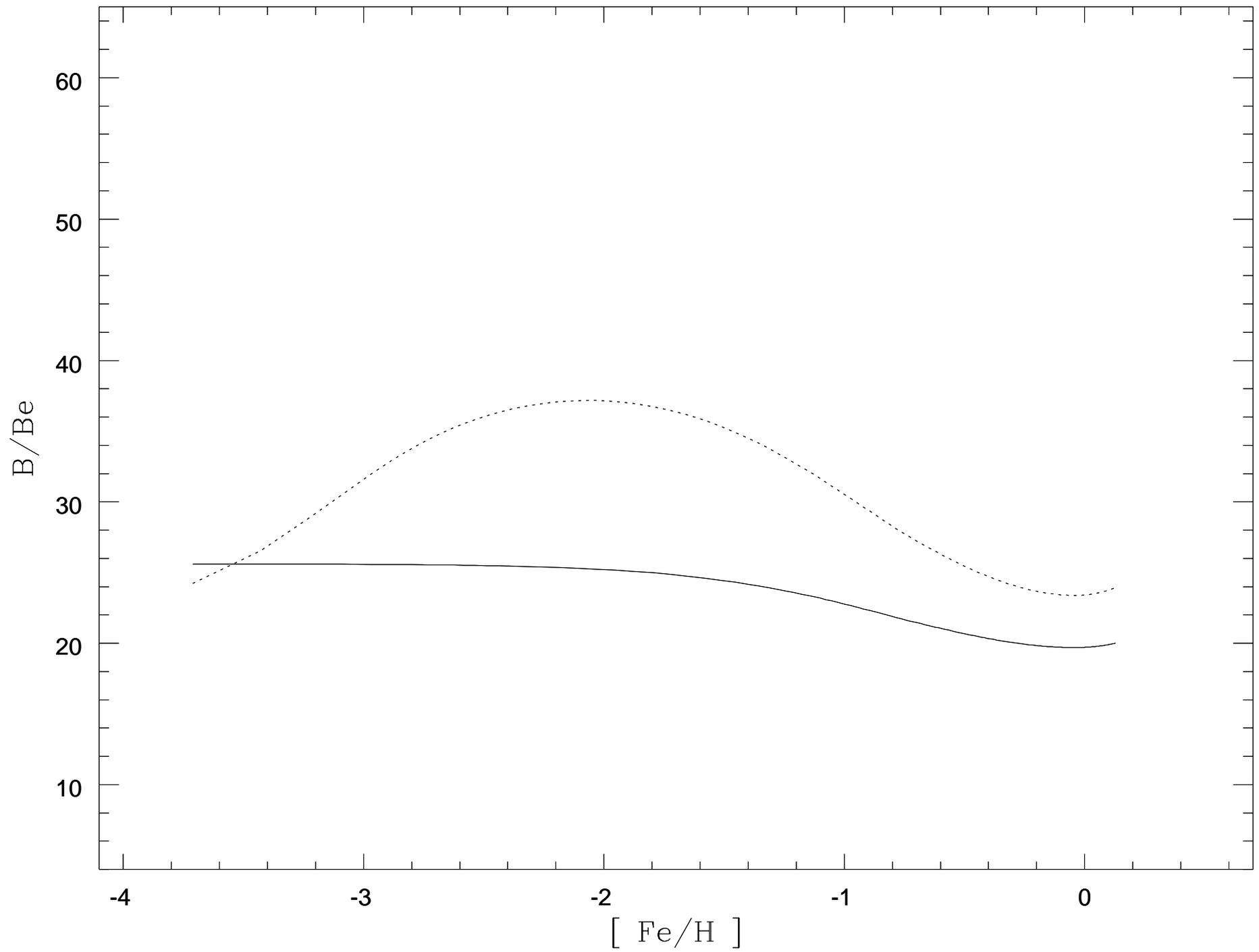
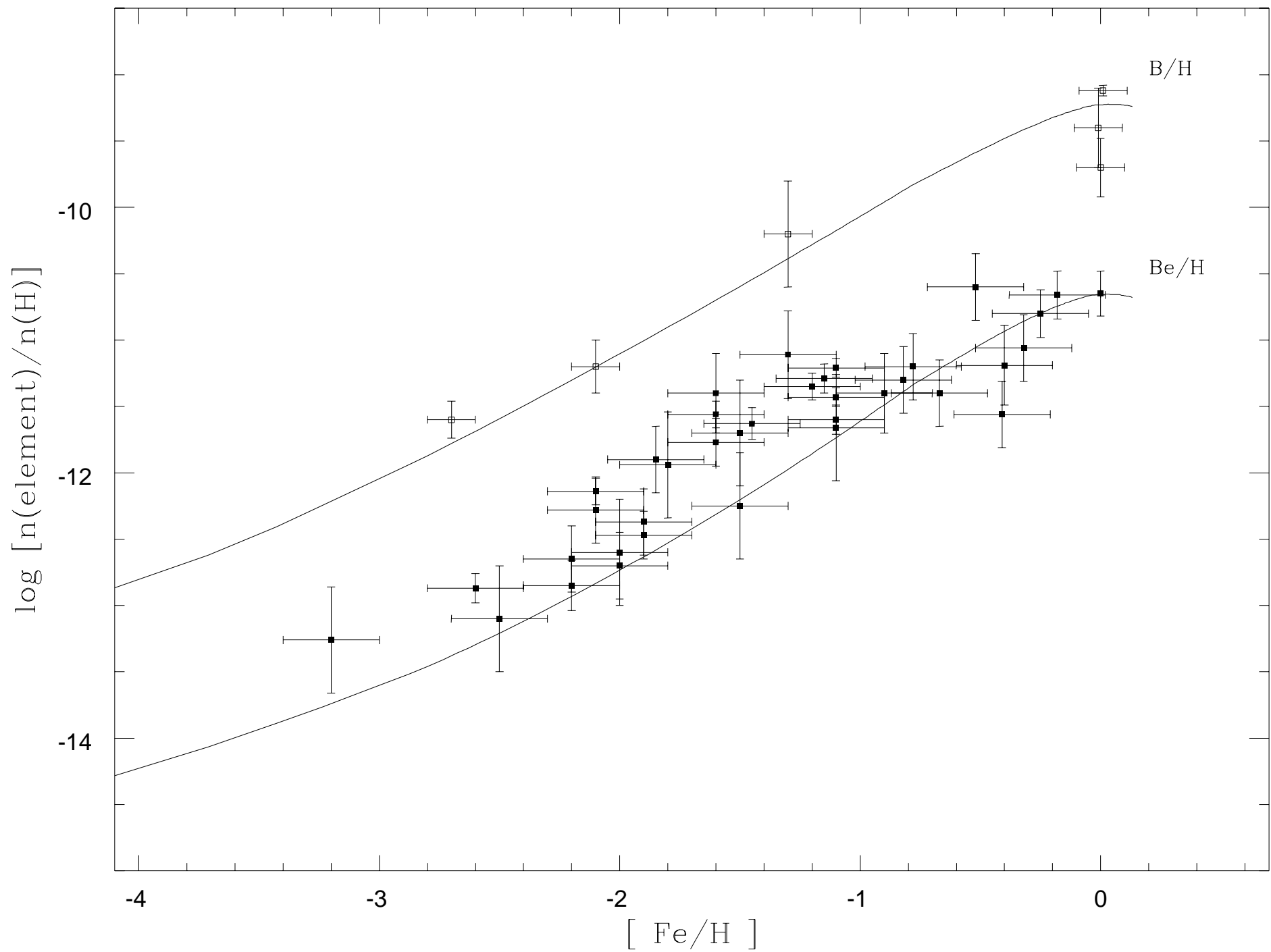
\end{document}